\documentclass[twocolumn,showpacs,prl,amsmath,amssymb]{revtex4}
\usepackage{graphicx}
\usepackage{bm}
\begin{document}

\title{Pseudospin Quantum Computation in Semiconductor Nanostructures}
\author{V. W. Scarola, K. Park, and S. Das Sarma}
\affiliation{Condensed Matter Theory Center, 
Department of Physics, University of Maryland,
College Park, MD 20742-4111}

\begin{abstract}
We theoretically show that spontaneously interlayer-coherent 
bilayer quantum Hall droplets should allow robust and 
fault-tolerant pseudospin quantum computation in semiconductor 
nanostructures with voltage-tuned external gates providing qubit 
control and a quantum Ising Hamiltonian providing qubit entanglement.  
Using a spin-boson model we estimate decoherence to be small 
$(\sim 10^{-5})$.  
\end{abstract}
\maketitle

Among the stringent requirements for viable quantum computer 
architectures are robust (i.e. relatively decoherence-free) and 
scalable qubits (i.e. quantum two-level systems) allowing single- and 
two-qubit operations necessary for quantum computation \cite{Divincenzo}.  
The scalability requirement makes semiconductor nanostructure based 
quantum computer architectures particularly attractive, and 
two spin-based semiconductor quantum computer architectures, one 
using electron spins in GaAs quantum dots \cite{Loss-Divincenzo} 
and the other using Si donor spin states \cite{Kane}, have attracted 
considerable attention.  The proposed advantage of solid state 
spin quantum computation over the corresponding charge or orbital 
state quantum computation is the long decoherence time for spin 
states ($\mu s$ or longer at low temperatures) compared with orbital 
states ($ps$ or less) allowing, at least in principle, robust 
quantum computation using spin qubits in semiconductor nanostructures.  
A very serious problem in solid state spin quantum computation is, however, 
the measurement of single spin states crucial for the 
quantum computation read-out.  
(There is no known solid state experimental technique for measuring 
a single spin, i.e. one Bohr magneton, and a great deal of experimental 
activity is currently being focused on measuring a single spin 
in semiconductor structures.)  In this letter we theoretically 
establish the practical possibility of a novel pseudospin 
quantum computation in semiconductor nanostructures which synergetically 
combines the robustness of spins (i.e. long decoherence time) with 
the ease of qubit-specific measurement of charge states by using 
mesoscopic ``coherent'' charge states in quantum Hall droplets 
\cite{MacDonald,evenodd}.  

Quantum Hall systems offer well studied mesoscopic quantum states     
with the potential for dynamic manipulation with long 
dephasing times.  Surprisingly little work has gone into exploring 
the possibility of engineering quantum Hall states for the purpose of 
quantum computation.  Mozyrsky \emph{et al.} \cite{Privman} 
have explored the possibility 
of using nuclear spins as qubits with an interaction mediated by 
a two-dimensional electron gas in the quantum Hall regime.  
Recently, Yang. \emph{et al.} \cite{MacDonald} have made a proposal for a 
quantum Hall two-level system using the charge degrees of freedom 
in two vertically coupled quantum
dots in a large magnetic field, a system which is currently the
subject of intense experimental study ~\cite{Tarucha}.  
In these systems the layer degree 
of freedom acts as a pseudospin, controllable through external
gates.  
The incompressibility of the finite size quantum Hall liquid
preserves the integrity of the two-level system while   
the mapping between layer index and pseudospin relies
on the presence of spontaneous interlayer phase coherence \cite{Eisenstein}.  
Drawing upon the
direct analogy between number fluctuations in the Cooper pair box
experiment ~\cite{Nakamura} and fluctuations in the layer 
degree of freedom in bilayer
quantum Hall droplets (BQHDs) an even-odd effect in the Coulomb
blockade spectra of BQHDs has been proposed \cite{evenodd} 
as a simple measure of
spontaneous interlayer phase coherence (and hence the robustness) of the 
two-level system discussed in Ref.~\cite{evenodd}.  

The remainder of this article will be concerned with entangling two
BQHDs which, when isolated, demonstrate the even-odd effect
independently.  Establishing controllable entanglement is 
crucial to performing large scale quantum computing.  
Our primary result is that the Coulomb interaction 
offers a natural entangling mechanism, opening the possibility of 
large scale quantum computing using BQHDS.  We find that, for
weak coupling, the Coulomb interaction between two laterally separated
BQHDs can be mapped
onto a quantum Ising model with a tunable, effective magnetic field.
This two-qubit Hamiltonian allows for relatively simple implementation of a
controlled-NOT operation \cite{Schon} which, when combined with single 
qubit operations, provides a universal set of quantum gates \cite{Barenco}. 
We further address the extent to which phonons and voltage fluctuations
in the leads dephase our system.


We begin by considering a set of two parabolically confined quantum
dots vertically separated by a distance $d$ under a transverse
magnetic field, $B$. 
The two dots will form a BQHD for appropriate magnetic fields
and layer spacings.
We further assume there to be a small, odd
number of electrons distributed between the two droplets.  In a large
magnetic field the Coulomb interaction exhibits energy cusps at
configurations corresponding to bulk, bilayer quantum Hall states.  We
focus our attention on the maximum density droplet (MDD) which is the 
mesoscopic realization of the bilayer quantum Hall state at 
total Landau level filling $\nu_T=1$. 

The Hamiltonian for an isolated BQHD in the Fock-Darwin basis is:
\begin{eqnarray}
H &=& H_0 + \hat{P} V_{\textrm{coul}} \hat{P}
\label{Hamiltonian}
\end{eqnarray}
where 
$H_0 = \frac{1}{2} 
\left( \sqrt{\omega^2_c +4\omega^2_0}
-\omega_c \right) \hat{L}_z $,
with $\hat{L}_z$ being the total angular momentum in the $z$-direction. 
Also, $\omega_c$ is the cyclotron frequency and 
$\omega_0$ parameterizes the parabolic confining potential.  
$\hat{P}$ is the lowest Landau level (LLL) projection operator and 
$V_{\textrm{coul}}$ represents the usual Coulomb interaction
between electrons:
\begin{eqnarray}
\frac{V_{\textrm{coul}}}{e^2/\varepsilon a} = \sum_{i<j \in \uparrow} \frac{1}{r_{ij}} 
+\sum_{k<l \in \downarrow} \frac{1}{r_{kl}}
+ \sum_{i\in \uparrow,k\in\downarrow} 
\frac{1}{\sqrt{r^2_{ik}+(\frac{d}{a})^2}},
\end{eqnarray}
where $\varepsilon$ is the 
GaAs dielectric constant,
and $r_{ij}$ is the lateral separation between
the $i$-th and $j$-th electron.  
The natural unit of length 
is the modified magnetic length 
$a=l_B (1+4 \omega_0^2/\omega_c^2)^{-1/4}$ which 
reduces to the planar 
magnetic length, $l_B =\sqrt{\hbar c/e B}$, when the 
cyclotron energy is much larger than the confining potential energy.
In the above we have used a 
pseudo-spin representation to describe
the double layer system: 
$\uparrow$ and $\downarrow$ distinguish different layers. 
In general we define the pseudo-spin operator:
\begin{eqnarray}
{\bm S} \equiv 
\frac{1}{2} \sum_{m} c^{\dagger}_{a}(m) \bm{\sigma}_{ab} c_{b}(m),
\end{eqnarray}
where $c^{\dagger}_{a} (c_{b})$ creates (annihilates) an electron in 
the layer $a(b)$ with single particle angular momentum $m$.  
${\bm \sigma}$ are the usual Pauli matrices.
$\hat{S}_z$ measures 
half the electron number difference between layers,
and $\hat{S}_x$ is associated with interlayer tunneling.  
We take the real spin to be
fully polarized 
either because of the large Zeeman coupling
or because of electron-electron repulsion, i.e. Hund's rule. 

We diagonalize $H$ in the 
basis of LLL single-particle eigenstates. 
In particular,
we focus our attention 
on the part of the Hilbert space containing the MDD state.  
It was shown in Ref.~\cite{evenodd} 
that, in the absence of tunneling, the two degenerate states 
with $S_z=\pm\frac{1}{2}$ are separated from states with different
$S_z$ by a large charging energy.  These two states, labeled
$\mid\uparrow\rangle$ and $\mid\downarrow\rangle$, form the qubit basis of 
our two-level pseudospin system.

\begin{figure}
\includegraphics[width=3.5in]{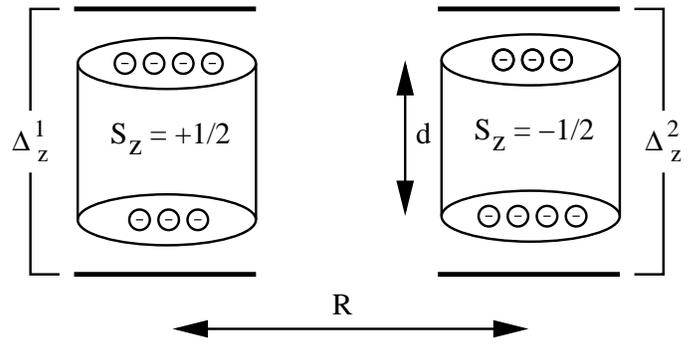}
\caption{Schematic representation of two bilayer quantum Hall 
droplets separated by a center-to-center distance $R$.  
Individual droplets are 
vertically separated by a distance $d$.  The left set of droplets has one
extra electron in the top layer giving it a net pseudospin,
$S_z=+1/2$.  The right set of droplets has pseudospin $S_z=-1/2$.   
This configuration corresponds to the basis state $\mid \uparrow \downarrow \rangle$.
$\Delta_z^1$ and $\Delta_z^2$ are the relative bias voltages 
between the layers in the left and right set of droplets, respectively.
\label{fig1}}
\end{figure}

Now consider a second BQHD along the $x$-axis at a distance 
$R$ away from the first 
BQHD, as shown in Fig.~\ref{fig1}.  To avoid lateral 
tunneling we keep the distance between BQHDs larger than 
the MDD diameter which is the diameter of the 
largest Fock-Darwin orbital, roughly $2\sqrt{N}a$, where $N$ is
the total number of electrons in both BQHDs.  
To include the inter-BQHD Coulomb interaction we first note that for $R\gg a$
the low lying energy levels of the two BQHDs contain a set of four
degenerate product states.  For $R\sim 8a$ the Coulomb interaction
between electrons in different BQHDs will favor the two 
pseudospin unpolarized states.  
We will verify, by direct calculation, that the inter-BQHD 
interaction energy can be made smaller than the intra-BQHD energy gap.  We
will consider a regime where the inter-BQHD interaction is 
too weak to produce 
intra-BQHD excitations, leaving the density unperturbed.   
Then, to a first approximation, we may take the 
basis states of the weakly interacting system of two BQHDs to be 
product states.  
 
We now calculate the inter-BQHD interaction matrix between the four
product states:  
\nolinebreak $\left\{\mid\uparrow\uparrow\rangle,\mid \uparrow\downarrow\rangle,\mid \downarrow\uparrow\rangle,\mid \downarrow\downarrow\rangle \right\}$.  
First note that 
the Coulomb interaction does not flip pseudospin so that  
all off-diagonal matrix elements vanish.   
The four diagonal matrix elements do not
vanish.  The inter-BQHD coulomb interaction, within our restricted 
Hilbert space, therefore maps onto a Ising interaction:
\begin{eqnarray}
H_I=\frac{J}{2}\sigma^1_z\sigma^2_z,
\end{eqnarray}
where we define the exchange splitting to be: 
\begin{eqnarray}
J&=&\langle \uparrow\uparrow\mid \sum_{i,j}V(R,d;{\bm r}_i,
{\bm r}_j^{\prime})\mid \uparrow\uparrow\rangle
\nonumber
\\ 
&-&\langle\uparrow\downarrow\mid \sum_{i,j}V(R,d;{\bm r}_i, {\bm r}_j^{\prime})\mid \uparrow\downarrow\rangle,
\end{eqnarray}
where:
\begin{eqnarray}
&V&(R,d;{\bm r}_i, {\bm r}_j^{\prime})=
\nonumber
\\
&&\frac{e^2}{\varepsilon ' a}
\frac{1}{\sqrt{(x_i-x_j'+R/a)^2+(y_i-y_j')^2+(d/a)^2}}.
\end{eqnarray}
${\bm r}$ $({\bm r}^{\prime})$ indicates the radial vector in the $x$-$y$
plane in the left (right) BQHD and $\varepsilon '$ is the
inter-BQHD dielectric constant.  
Note that with this definition the Coulomb interaction will favor 
an antiferromagnetic interaction with $J>0$.  The states $\mid \uparrow\rangle$ and 
$\mid \downarrow\rangle$ are nontrivial,
many-body eigenstates of $H$.  

\begin{figure}
\includegraphics[width=3.0in]{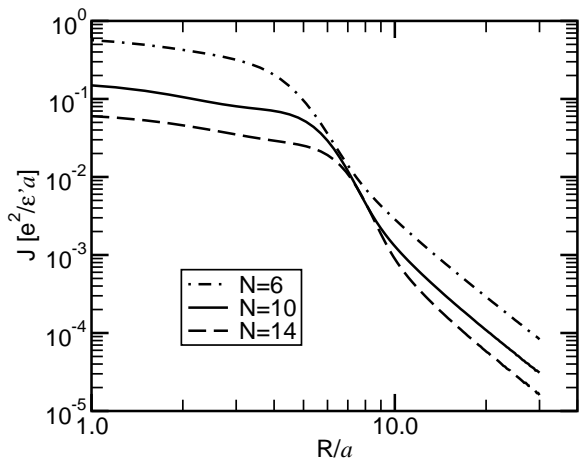}
\caption{
The exchange splitting, $J$, between the pseudospin 
singlet and triplet states of a pair of bilayer quantum Hall droplets
versus lateral separation $R$.  $J$ is evaluated 
for systems with a total of 6,10, and 14 electrons.  
The vertical separation between droplets is $d=a$.  
The droplet diameter is below $7.5 a$ for each curve.  $\varepsilon '$
is the dielectric constant between bilayer quantum Hall droplets.
\label{fig2}}
\end{figure}

Fig.~\ref{fig2} is a log-log plot of $J$ as a function 
of the inter-BQHD separation $R$ for a
total of 6,10, and 14 electrons distributed between two BQHDs with
vertical spacing $d=a$.  For $R\sim 8a$ we find
that $J$ is appropriately smaller than the intra-BQHD edge excitation
gap which is $\sim0.04 \frac{e^2}{\varepsilon a}$ \cite{evenodd}.  
At sufficiently large distances an odd number of
electrons in a single BQHD in the MDD state can be thought of as a
dipole.  Fig.~\ref{fig2} shows that $J\propto R^{-3}$ for $R\gtrsim 25a$, 
demonstrating that the inter-BQHD interaction is dipolar {\em only} for 
sufficiently large distances.  
In the regime of interest $R\sim8a$, $J$ decreases 
much faster than $R^{-3}$.  We find that for $R\sim20a$ the exchange 
splitting is already as low as $J\sim0.5 \mu$ eV at $B=9 T$.  
This suggests that the interaction 
between a collection of BQHDs will be effectively nearest neighbor 
given that we have not considered effects such as finite 
layer thickness and, in general, screening of the 
inter-BQHD interaction, which significantly 
reduce the strength of the Coulomb interaction. 

We now allow for interlayer tunneling within a single
BQHD.  The tunneling Hamiltonian can be written in terms of the 
pseudospin operator:
\begin{eqnarray}
H_t= -t\hat S_x,
\end{eqnarray}   
where $t$ is the single particle, interlayer tunneling gap.  In the
reduced Hilbert space, $\mid \uparrow \rangle$ and $\mid \downarrow\rangle $, of the $i$th
BQHD we find \cite{MacDonald,evenodd}:
\begin{eqnarray}
H_{\textrm{red}}^i = - \Delta_x^i \sigma_x^i +\Delta_z^i \sigma_z^i.
\end{eqnarray} 
In the limit of a small single-particle tunneling gap $t$,
$\Delta_x = t \langle \uparrow \mid \hat{S}_x \mid \downarrow \rangle$.
Also,
$\Delta_z$ is the relative bias voltage between layers.
$\Delta_x$ is the {\it renormalized tunneling gap}
which is greatly enhanced
from the single-particle tunneling gap, $t$,
by the Coulomb interaction.  $H_{\textrm{red}}$ acts as the Hamiltonian of an 
effective magnetic
field pointing in the $x$-$z$ plane.  The effective field will reorient the
direction of the on-site pseudospin.

The system discussed here has the advantage of being scalable.
One may consider a large number of BQHDs coupled 
via nearest neighbor
interactions.  Two examples include a 
linear chain of closely spaced
BQHDs or a planar, triangular lattice.   The reduced
Hamiltonian of a weakly coupled, many-BQHD system corresponds to a 
quantum  Ising model:
\begin{eqnarray}
H_{\textrm{total}} = \sum_{i}[ -\Delta_x^i \sigma_x^i +\Delta_z^i \sigma_z^i]
+ \sum_{i,j}\frac{J^{ij}}{2} \sigma_z^i \sigma_z^j.
\end{eqnarray}
To perform a quantum logic operation the single qubit parameters in 
$H_{\textrm{total}}$ should be tunable.  First, $\Delta_z$ may be   
adjusted by applying a gating bias to each BQHD.  
$\Delta_x$ can be tuned by changing $t$ through an in-plane 
magnetic field or a gating mechanism which alters the lateral 
position of the dots.
The inter-qubit parameter, $J$, may be tuned by 
placing a third BQHD between the original two BQHDs.  The inter-BQHD 
interaction can be turned on and off by placing an even or 
odd number of 
electrons in the intermediate BQHD or by depleting it completely.  
In fact it is not necessary to physically change $J$ because 
it may be possible to effectively tune the qubit coupling 
through a series of refocusing pulses.
This technique has been used to implement 
quantum algorithms in NMR liquids governed by $H_{\textrm{total}}$,  
where the fixed coupling is between nuclear spins.  
It is also important to note that the architecture proposed 
here has the additional advantage of being charge based, simplifying 
read out.  
Single electron transistors, in principle, already have
the capability \cite{Schon,Schoelkopf} 
of measuring the charge imbalance between 
states with $S_z=+1/2$ and $S_z=-1/2$.

\begin{figure}
\includegraphics[width=2.3in]{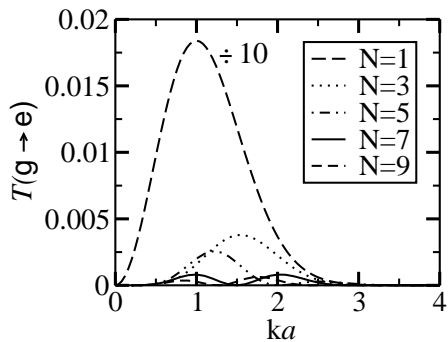}
\caption{
The probability, $T$, that a perturbation of the electron density 
will excite a bilayer quantum Hall droplet from the 
ground, maximum density droplet state to an edge state as a function 
of wave vector, $k$.  
The vertical separation between droplets is $d=a$.  
The particle number, $N$, is increased from 1 to 9, showing a dramatic
decrease in $T$.
\label{fig3}}
\end{figure}

Finally, we consider the important issue of qubit robustness by 
showing long pseudospin coherence times in BQHD systems.  
We consider two sources of dephasing in a single BQHD: phonons and 
voltage fluctuations in the leads.  Phonons readily couple to 
single electron degrees of freedom in quantum dots.  This 
may potentially destroy our proposed two level system through 
leakage to excited states.  
We note, however, that the rigidity of the  
incompressible quantum Hall droplet lifts the large degeneracy 
of the excited states, thereby suppressing phonon induced excitations.  
To show this quantitatively, consider the following general
Hamiltonian for electron-phonon coupling:
\begin{eqnarray}
H_{\textrm{e-p}}=\sum_{{\bm k}}M_{{\bm k}}\rho({{\bm
    k}})(a^{\vphantom{\dagger}}_{{\bm k}}
+a^{\dagger}_{-{\bm k}}),
\end{eqnarray}
where $M$ is an arbitrary electron-phonon interaction matrix 
element and $a^{\vphantom{\dagger}}_{{\bm k}}$ 
and $a^{\dagger}_{{\bm k}}$ annihilate and 
create phonons of wave vector ${\bm k}$ in the $x-y$ plane.
$\rho({\bm k})$ is the density operator, given by:
\begin{eqnarray}
\rho({\bm k})=\sum_{j=1}^{N}e^{i{\bm k}\cdot{{\bm r}_j}} .
\end{eqnarray}
We use first order perturbation theory to estimate 
the change in the rate at which phonons couple to our two level system as we 
increase the system size.  In which case the 
electron-phonon scattering rate is proportional to the 
transition matrix element between the initial and final electronic states.  
We calculate the transition matrix element between 
our proposed two-level ground state, $g$, and the lowest excitation, $e$:
\begin{eqnarray}
T(g \rightarrow e)=\vert\langle e \vert
\rho({\bm k})
\vert g\rangle \vert^2.   
\end{eqnarray}
$T$ measures the probability that    
a phonon of wave vector $\bm{k}$ will 
induce an excitation from the ground state to the excited state. 
The ground and excited states are computed from $H$ using 
exact diagonalization at angular momenta corresponding to 
the MDD and its edge excitation, respectively \cite{evenodd}.  
Fig.~\ref{fig3} plots the matrix element $T$ as a function 
of $\vert\bm{k}\vert$ for several different particle numbers.  From the 
plot we see that 
phonon coupling to edge modes is suppressed as we increase the system 
size suggesting that the incompressible system studied 
here will be less sensitive to dephasing from phonons than 
analogous systems utilizing single particle, charge degrees of freedom.

Another primary mechanism coupling pseudospin to the environment is 
similar to the corresponding Cooper pair box problem \cite{Schon}, 
through voltage fluctuations, $\delta V$, (e.g. in the gating potential) 
leading to the standard spin-boson model for decoherence of 
a single qubit:
\begin{eqnarray}
H_{\textrm{SB}} = - \Delta_x \sigma_x +\Delta_z \sigma_z+\gamma e\delta 
V\sigma_z,      
\end{eqnarray}
where $\gamma$ is a dimensionless parameter related to the qubit and 
gate capacitances.  Following standard spin-boson techniques \cite{Schon}, 
the voltage fluctuations due to an external impedance with Ohmic
dissipation can be 
modeled by a harmonic oscillator bath, leading to an equivalent mapping 
of our pseudospin decoherence problem to the corresponding Cooper pair box 
decoherence \cite{Nakamura,Schon} problem.  This then leads, after some 
straightforward algebra, to the following low temperature 
estimate for the decoherence factor $\alpha$, which is the ratio 
of the dephasing rate to the elementary logic operation rate:
\begin{eqnarray}
\alpha=\gamma^2\frac{4R_v}{R_K}
=\left[\frac{C_g}{2C_g+C_d}\right]^2 \frac{4R_v}{R_K},
\end{eqnarray}
where $C_g$ and $C_d$ are the gate and quantum dot capacitances, 
respectively;  $R_v (\sim50\Omega)$ is the typical impedance of the 
voltage circuits, and $R_K =h/e^2$.  Using results from Ref.~\cite{evenodd} 
to estimate the dot capacitance and reasonable values 
for the gate capacitance we get $\alpha\sim10^{-5}$, establishing that 
robust fault-tolerant quantum computation should be possible in pseudospin 
quantum Hall systems.  Although voltage fluctuations are likely 
to be the dominant decoherence mechanism on our proposed BQHD qubit, there 
are other possible dephasing channels which should be considered in
the future.  In particular, we suggest that the time scale 
of $1/f$ noise associated with charge fluctuations \cite{Nakamura2} 
is long enough to be dealt with using refocusing.   

We acknowledge helpful conversations with J.K. Jain.  This work 
was supported by ARDA.



\end{document}